\documentclass[english,10pt]{article}
\usepackage{amsmath}

\usepackage{geometry}
\usepackage{times}
\usepackage{bm}
\usepackage{natbib}

\usepackage[plain,noend]{algorithm2e}

\makeatletter
\renewcommand{\algocf@captiontext}[2]{#1\algocf@typo. \AlCapFnt{}#2} 
\def\@algocf@capt@plain{top}
\renewcommand{\algocf@makecaption}[2]{%
  \addtolength{\hsize}{\algomargin}%
  \sbox\@tempboxa{\algocf@captiontext{#1}{#2}}%
  \ifdim\wd\@tempboxa >\hsize
    \hskip .5\algomargin%
    \parbox[t]{\hsize}{\algocf@captiontext{#1}{#2}}
  \else%
    \global\@minipagefalse%
    \hbox to\hsize{\box\@tempboxa}
  \fi%
  \addtolength{\hsize}{-\algomargin}%
}
\makeatother

\newcommand\independent{\protect\mathpalette{\protect\independenT}{\perp}}
\def\independenT#1#2{\mathrel{\rlap{$#1#2$}\mkern2mu{#1#2}}}

\usepackage{csquotes}

\usepackage{etoolbox}
\BeforeBeginEnvironment{longtable}{\begin{center}\small}
\AfterEndEnvironment{longtable}{\end{center}}

\usepackage{tikz}
\usepackage{subcaption}
\usepackage{mathtools}

\usepackage{amsfonts}

\newcommand{\eqd}{\stackrel{\mathrm{d}}{=}}

\usepackage{graphicx}
\newcommand{\indep}{\rotatebox[origin=c]{90}{$\models$}}
\usepackage{url}

\definecolor{ejc}{RGB}{255,0,0}
\definecolor{ms}{RGB}{0,125,0}
\definecolor{yxc}{RGB}{0,0,200}

\usepackage{parskip} 

\begin{document}

\title{Rejoinder: \\``Gene Hunting with Hidden Markov Model Knockoffs''}

\date{December 10, 2018}

\author{M. SESIA, C. SABATTI, E. J. CAND\`ES}
\author{Matteo Sesia\thanks{Department of Statistics, Stanford
 University, Stanford, CA 94305, U.S.A.}
\and
Chiara Sabatti\footnotemark[1] \thanks{Department of Biomedical Data Science, Stanford
 University, Stanford, CA 94305, U.S.A.}
\and
Emmanuel J. Cand\`es\footnotemark[1] \thanks{Department of Mathematics, Stanford
 University, Stanford, CA 94305, U.S.A.}}

\maketitle

\begin{abstract}
In this paper we deepen and enlarge the reflection on the possible advantages of a knockoff
approach to genome wide association studies \citep{sesia2018}, starting from the discussions in
\cite{bottolo2019comment, jewell2019comment, rosenblatt2019comment} and \cite{marchini2019comment}.
 The discussants bring up
a number of important points, either related to the knockoffs
methodology in general, or to its specific application to genetic
studies. In the following we offer some clarifications, mention
relevant recent developments and highlight some of the still open
problems.
\end{abstract}

\section{Conditional vs. marginal hypotheses}
The model-X framework of knockoffs \citep{candes2016} addresses a
general multiple testing problem in which the $j$th null hypothesis
states that the response $Y$ is independent of the explanatory
variable $X_j$ conditional on all other predictors $X_{-j}$.  In the
very special case where the joint distribution of $X$ and $Y$ is
multivariate Gaussian, this is equivalent to testing for the presence
of conditional or partial correlations \citep{rosenblatt2019comment}.
In general, the non-null variables in the model-X framework are those
belonging to the Markov blanket of $X$ on $Y$
\citep{candes2016}.

There is little sense in arguing that, generally speaking, conditional
or marginal hypotheses are the ``right ones'' to test: the choice
between these two approaches will clearly depend on the problem at
hand. For example, if $Y$ describes the cancer status of a tissue
sample, and each $X_j$ the expression level of gene $j$, one might be
interested in testing the hypotheses that $Y\independent X_j$. The
rejections of these null hypotheses would describe all genes whose
expression level changes with cancer status.  Whenever we think of
relating a response $Y$ with a linear (or generalized linear) model in
term of $X$, and test the hypotheses that the regression coefficients
vanish, we are testing hypotheses that are of a conditional flavor
(these hypotheses correspond exactly to the conditional ones in the
case where the response follows a generalized linear model, see
\cite{candes2016}).

In a genome-wide association study, we find ourselves precisely in a
situation of the latter type. The traits of interest are 
polygenic---they are influenced by the contribution of many genetic variants---and
the models most commonly used by the scientific community are either
linear or log-linear. Testing conditional hypotheses corresponds to
trying to identify those genetic variants whose coefficients are
nonzero in these models.  As we pointed out in our paper, the
literature has attempted to apply multivariate models since the very beginning;
see for example \cite{Hoggart2008}. While the scientific interest of the
conditional hypotheses was never in question, the genetic community
encountered a number of challenges in the application of multivariate
methods that only recently we have begun to be able to address. A
fundamental difficulty has been the inability to couple the findings
of procedures such as the Lasso with precise reproducibility
guarantees: in order to be able to associate a p-value to each of the
genetic variants in the study, researchers resorted to marginal
analysis. The knockoffs approach, by guaranteeing control of the false 
discovery rate over the selected variants, bypasses this difficulty and therefore opens the
possibility of analyzing the data with those models that geneticists
always thought provided a more accurate description of reality.

Rather than repeating ourselves, we would like to take the occasion to
augment the discussion of this topic with some additional references.
Firstly, let us point out that, as underscored by
\cite{marchini2019comment}, the standard methods of analysis of genome-wide association 
data already depart from an entirely marginal framework. By relying on
linear mixed models \citep{zhang2010mixed,KetE10} with a covariance
matrix estimated from the entire genotype data, geneticists
effectively try to estimate the contribution of a specific variant
$X_j$ in addition to that of the rest of the genome $X_{-j}.$ While
this approach has proven a step forward, it still suffers from some
important limitations: for example, it relies on fairly restrictive
distributional assumptions, and requires the researcher to postulate a
different model relating the phenotype $Y$ to the genotype $X$ for
every variable that is analyzed. Possibly one of its most important limits
is that it is unable to resolve the contribution of multiple variants
in linkage disequilibrium---a topic that we shall discuss in greater depth
later. 

Secondly, we would like to refer the reader to two recent
contributions \citep{BetB16,KetB16} to the literature of 
genome-wide association studies going in a
direction similar to ours; that is to say, enabling a multivariate
analysis with some reproducibility guarantees. While these authors attempt to control the 
family-wise error rate,
\cite{BetB16} in particular has interesting remarks on the conditional
versus marginal hypotheses: they observe how, under appropriate
assumptions, if ``the coefficient [for the single nucleotide polymorphism (SNP) $j$] in the multivariate
linear regression is different from zero,
[...]  
there exists a non-zero direct causal effect from SNP $j$ to the
phenotype $Y$. This statement is not true with marginal associations
(i.e. if SNP $j$ is only marginally associated with $Y$) since
adjusting for all other SNPs (different from SNP $j$) is crucial for
causal statements.''  Since the ultimate scientific goal is to
identify causal variants \citep{edwards2013beyond, visscher201710},
this is a strong argument in favor of conditional hypotheses.

Finally, we want to underscore another sense in which marginal hypotheses are becoming less interesting in genome-wide association studies. As sample sizes $n$ increase, it has become apparent that polygenic traits are indeed influenced by a very large number of genetic variants \citep{boyle2017expanded}. Coupling this with the presence of linkage disequilibrium and the fact that large $n$ allows one to detect even small departures from independence, one realizes that, soon enough, we will be in the position to reject the marginal null $Y\independent X_j$ for every $X_j$ in the genome, an utterly uninteresting result. 

\section{False discovery rate vs. family-wise error rate}

Even though the family-wise error rate is still the most commonly used
measure of type-I errors in genome-wide association studies \citep{rosenblatt2019comment}, we
feel quite strongly that the false discovery rate is arguably
more appropriate. As modern studies of complex traits often lead to
the discovery of several hundreds of loci \citep{visscher201710,
  boyle2017expanded}, it seems excessive to worry about the
probability of reporting a single false finding, especially when large
leaps of faith are involved in the traditional postulation of the null
hypotheses and in the assumptions of the linear models. The
statistical genetics community has already widely accepted the concept
of false discovery rate for the analysis of gene expression and other
genomic measurements \citep{GTEx}. It is more plausible that its
adoption in genome-wide association studies has been hindered by the methodological difficulties
arising from the correlations among the variants, rather than any
fundamental objections to its principle. Therefore, as new statistical
methods are developed, we expect that the use of the false discovery rate 
in genome-wide association studies will keep on expanding.

\section{The resolution of conditional testing}
In a genome-wide association study, each explanatory variable $X_j$ can naturally be chosen to
represent a single nucleotide polymorphism, so that feature
selection will be performed at the highest possible resolution allowed
by the genotyped data. However, unless the signals are sufficiently
strong, conditional testing may be a hopeless task and different
hypotheses should be analyzed instead. For example, if $X_1$ and $X_2$
are nearly identical within the collected sample, it may be wiser to
ask whether $Y \indep (X_1,X_2) \mid X_{-\{1,2\}}$ rather than
$Y \indep X_1 \mid X_2, X_{-\{1,2\}}$ and
$Y \indep X_2 \mid X_1, X_{-\{1,2\}}$. Consequently, if knockoffs are
applied to the individual hypotheses, $\tilde{X}_1$ and $\tilde{X}_2$
will certainly almost be equal to $X_1$ and $X_2$, and thus powerless
\citep{rosenblatt2019comment}. It is important to underline that this
is not a limitation of our method. Instead, it is an inevitable
reflection of the fundamental undecidability of the question that was
asked, as conditional testing may only be performed at the resolution
allowed by the data.

The solution adopted in our paper is that suggested in \cite{candes2016}: the variants are grouped based on their empirical correlations and knockoffs are only constructed for a set of promising prototypes identified through a suitable data carving scheme. Even though the choice of a resolution may be somewhat arbitrary in our paper, our methods can be easily applied with different values of the clumping correlation threshold. It is left to future research to determine whether an optimal choice exists, and how to combine the results obtained at different resolutions. As correctly pointed out in \cite{jewell2019comment}, our approach formally amounts to asking whether $Y \indep X^*_j \mid X^*_{-j}$, where $X^*_j$ indicates the prototype for the $j$th group. 

Alternatively, one could directly test group-wise hypotheses of the type $Y \indep (X_{1}, X_2) \mid X_{-\{1,2\}}$ by extending the notion of group knockoffs \citep{Dai2016, KS18} to our methods. This approach arguably offers a more elegant interpretation as it completely avoids the pruning of any markers \citep{marchini2019comment}, at some additional computational cost. For this purpose, we have already developed new efficient algorithms that will soon be presented as part of our follow-up work.

To put this in the context of the genetics literature, we note how the
standard analysis of genome-wide association studies allows only the identification of loci that are
(marginally) associated with the trait of interest, without
discriminating between the many variants that are present at these
loci. To increase the resolution of the findings, one resorts to what
are known as ``fine-mapping'' methods
\citep{hormozdiari2014identifying,spain2015strategies}. These
invariably rely on a multivariate model, and are also faced with the
impossibility of resolving the signal beyond the level of information
present in the data. For example, \cite{hormozdiari2014identifying}
output a ``causal set'' of variants that is guaranteed to contain the
truly causal ones, but will also include others, practically
indistinguishable from these. An interesting feature of the
knockoff-based approach to genome-wide association studies is that, effectively, it performs
simultaneously locus identification and fine mapping.

\section{Confounders}

The confounding effect of an inhomogeneous population is a major
source of concern in the {\em marginal} analysis of genome-wide
association studies \citep{pritchard2000inference} and it is not
surprising that the discussants bring this up
\citep{marchini2019comment, bottolo2019comment}. Because we do not
expect the wide readership to be familiar with this
issue, it is best to explain it in a few lines through a stylized 
example. Imagine that the statistician has
available genotypes of individuals blindly sampled from the European and
African populations, which have substantially different diets.
Suppose further that our statistician is
interested in the genetic determinants of blood lipid levels, which are also influenced by dietary intakes, and hence have a different mean level in the European and African population.  Then all the
variants $X_j$ which differ in frequency across the two populations (and there are many)
will show a strong association with the response $Y$ and may get
picked up. However, they may have no direct genetic link to blood lipids. 
Rather, a strong signal may be observed simply because the value of a marker is correlated
to the population an individual belongs to, and  different
populations have different diets, and that diets influence blood lipid
levels.

As recognized before \citep{KetB16}, conditional
testing already implicitly accounts for any population structure. To
quote from \cite{KetB16}, ``testing of markers with a high-dimensional
variable selection procedure, which can account for the correlations
between the markers, does not require any population structure
correction at all.''  This is simply because we are asking whether a
particular variant $X_j$ provides information about the phenotype {\em in
  addition to anything that can already be inferred} from the value
$X_{-j}$ of all the other hundreds of thousands of variables. Conditioning on
$X_{-j}$ implies conditioning on the different ancestries of the
individuals. To return to our example, if we get to see hundreds of thousands of
genetic variants about an individual, then we already know which
population this individual belongs to. In our analysis of the
Northern Finland 1966 Birth Cohort study, the first 5 principal
components of the genotype matrix are therefore included mainly to
increase power.

The real issue here concerns the validity of the sampling mechanism.
As observed in \cite{bottolo2019comment}, the hidden Markov model of
\cite{Scheet2006} is better suited to describe populations that are
homogeneous and unrelated, or that contain known patterns. If the
structure of the sub-populations is unknown, more complex models
\citep{Falush2003, delaneau2012linear, o2014general, o2016haplotype}
should be used to generate knockoffs. Since the modeling of
inhomogeneous populations typically relies on more refined hidden
Markov models, we can say in response to \cite{bottolo2019comment}
that the extension of our work is, at least conceptually, rather
straightforward. This merely involves novel computational challenges
that will be addressed in future work. Meanwhile, our current approach
can be further justified by observing that knockoffs tend to be quite
robust to some degree of model misspecification \citep{candes2016,
  barber2018robust,romano2018}.

\section{Case-control studies}

\cite{marchini2019comment} asks about the impact of the artificial
inflation in the frequency of the haplotypes surrounding the causal
variants in case-control studies, and we pause to discuss this
carefully. There are two populations we may want to think about: a {\em
  prospective population} $G_{XY}$ of individuals obeying certain
characteristics; for instance, all adult males living in the UK; and
a {\em retrospective population} $F_{XY}$ in which, to quote from
Marchini, cases are usually more prevalent. In the retrospective
population, the proportion of cases versus controls takes on an
arbitrary value, which is typically higher than that in the
prospective population. Formally, the relationship between the
prospective and retrospective populations is as follows:
\[
G_{X|Y} = F_{X|Y}, \quad G_Y \neq F_Y.
\]
The equality follows directly from the assumption that cases and
controls in the study are randomly sampled from the prospective
population of diseased and healthy individuals. The inequality is due
to the fact that the proportions of cases typically differ.
Consequently, the marginals are different as well, i.e.,
$G_X \neq F_X$.

In our work, we get independent samples from the retrospective
population $F_{XY}$. Thus, as long as the hidden Markov model provides
a good approximation for the marginal $F_X$, the method applies and
the inference is valid.  Since we estimate $F_X$ by relying on the
genotypes of the cases and controls contained in our sample, we
control the false discovery rate for testing
$Y \independent X_j | X_{-j}$ whenever $(X,Y) \sim F$.  Although we do
not prove this here, the definition of a null does not change whether
$(X,Y) \sim G$ or $(X,Y) \sim F$ (we can loosely say that the
definition of conditional independence does not depend on the
distribution of the covariates). We believe this answers Marchini's
comment on type-1 error control.

There is a broader issue of interest as well. To be sure, we often
claim that one attractive feature of the knockoffs approach is that we
may want to use lots of unlabeled data to `learn' the distribution of
the covariates $X$. If this were the case, we would learn $G_X$, not
$F_X$! We would then construct features that are exchangeable when
$X \sim G_X$, perhaps not when $X \sim F_X$.  {What is the implication
  of this?  A cool result is that despite this apparent mismatch,
  knockoffs constructed in this way provide valid inference as well
  when $X \sim F_X$!} This fact will be rigorously established in a
future publication. Our intent here is merely to explain that our
approach allows considerable flexibility in the way we construct
knockoff variables in case-control studies.

\cite{marchini2019comment} also asks about power. In light of our
discussion, we may ask whether we should build knockoffs based upon
$G_X$ or upon $F_X$ as to maximize power. This is an interesting but
delicate question, which requires more analysis than we can possibly
offer here.

\section{What makes good negative control variables?}

The idea to use pseudo-variables to guide the selection of important 
features is not new and goes back
at least to \cite{miller1984selection}, as remarked in
\cite{barber2015}. Having said this, there is a profound distinction
between a vague program and an operational procedure that achieves
clearly stated goals. This is best explained by focusing our remarks
on the work of \cite{wu2007}, which is brought up by
\cite{rosenblatt2019comment}, and on permutation techniques discussed
in \cite{bottolo2019comment}.

\cite{wu2007} state that ideal pseudo or phony variables should obey
two properties recalled in \cite{rosenblatt2019comment}: ``(A1) real
unimportant variables and phony unimportant variables have the same
probability of being selected on average'', and ``(A2) real important
variables have the same probability of being selected whether or not
phony variables are present''. This is a wishful list that their paper
does not show how to implement. In contrast,
the knockoffs framework gives us (1) some precise rules for
constructing synthetic features which can be safely used as negative
controls and (2) a concrete selection procedure---a filter---which
sifts through variable and knockoff scores computed via any method the
statistician suspects to be powerful while rigorously controlling the
false discovery rate \citep{barber2015}; this filter is unlike
anything we have seen in the literature. We expand on these two
novelties below.

Knockoff variables are entirely different from existing
pseudo-variables, including variables obtained by permutations, and we
make this clear through the simplest possible example. Imagine we have
$n$ i.i.d.~samples $(X_1^{(i)}, X_2^{(i)}, Y^{(i)})$,
$i = 1, \ldots, n$, drawn from a population in which
\[
(X_1, X_2) \sim \mathcal{N}(0, \Sigma) \quad \text{and} \quad Y | X_1,
X_2 \sim \mathcal{N}(X_1, 1),
\]
so that the first variable belongs to the linear model while the
second does not.  We assume that $X_1$ and $X_2$ have unit variance
and that $\operatorname{corr}(X_1, X_2) = 1/2$. By definition,
knockoff variables obey
$(\tilde X_1, \tilde X_2) \eqd ({X}_1, \tilde X_2) \eqd (X_1, X_2)$, so
that a knockoff feature correlates with a true feature in exactly the
same way as a pair of true features;
$\operatorname{corr}(X_1, \tilde{X}_2) = 1/2$. How do the four
pseudo-variable proposals of \cite{wu2007} compare?  In the first case, the
pseudo variables $X_1^*$ and $X_2^*$ are independent standard normal
and independent of anything else (the proposal would be the same if
the covariates are not Gaussian as long as each marginal has mean zero
and variance one).  Clearly, $(X_1^*, X_2^*)$ is not at all
distributed as $(X_1, X_2)$ and, moreover,
$\operatorname{corr}(X_1, X^*_2) = 0$.

In the second proposal, the pseudo-variables $(X_1^*, X_2^*)$ are
obtained by applying a random permutation, see also
\cite{bottolo2019comment}: concretely, the pseudo-variables for the
$i$th observation are $\{(X_1^{(\pi(i))}, X_2^{(\pi(i))})\}$, where
$\pi$ is a random permutation from $\{1, \ldots, n\}$. By
construction, we now have $(X_1^*, X_2^*) \eqd (X_1, X_2)$. However, a
simple calculation shows that whereas
\[
\operatorname{corr}(X_1, \tilde{X}_2) = \frac{1}{2}, \quad
\operatorname{corr}(X_1,  X_2^*) = \frac{1}{2n}.
\]
In the limit of large samples, the correlation between $X_1$ and
$X_2^*$ vanishes.  So it is, once again, completely
different. The remaining proposals in \cite{wu2007} are
  refinements of the first two and operate by projecting the pseudo
  variables above onto the orthogonal complement of the space spanned
  by the original covariates.

  Consider now what happens when we compute statistics for testing
  whether variables are in the model or not. Here, the sample
  correlation between $X_2$ and $Y$ has mean $1/2$ whereas that
  between $X_2^*$ and $Y$ vanishes (it is equal to $0.5/n$). Hence,
  the permutation $X_2^*$ cannot serve in any way as a negative
  control. To serve as a negative control, a phony variable needs to
  have the same explanatory power than the null variable being tested;
  colloquially, we might say that it needs to have the same $R^2$. A
  phony variable generated by a random permutation, however, is
  essentially independent of the response and has, therefore, no
  explanatory power whatsoever. These facts also apply to the
    forward selection method of \cite{wu2007} as it is easy to imagine
    examples in which true nulls have a much higher chance of being
    selected than permuted features.  In summary, permutation methods
  may be useful to test the existence of any relationship between a
  response and a family of covariates but they generally cannot be
  used to provide any finer-grained information
  \citep{DiCiccioRomano}.  It is, therefore, impossible to understand
  how the insights of \cite{wu2007} ``will later be formalized by
  knockoffs'' as suggested by \cite{rosenblatt2019comment}.

\begin{figure}[!htb]
  \centering
  \begin{subfigure}[!htb]{0.9\textwidth}
    \centering
    \includegraphics[width=0.8\textwidth]{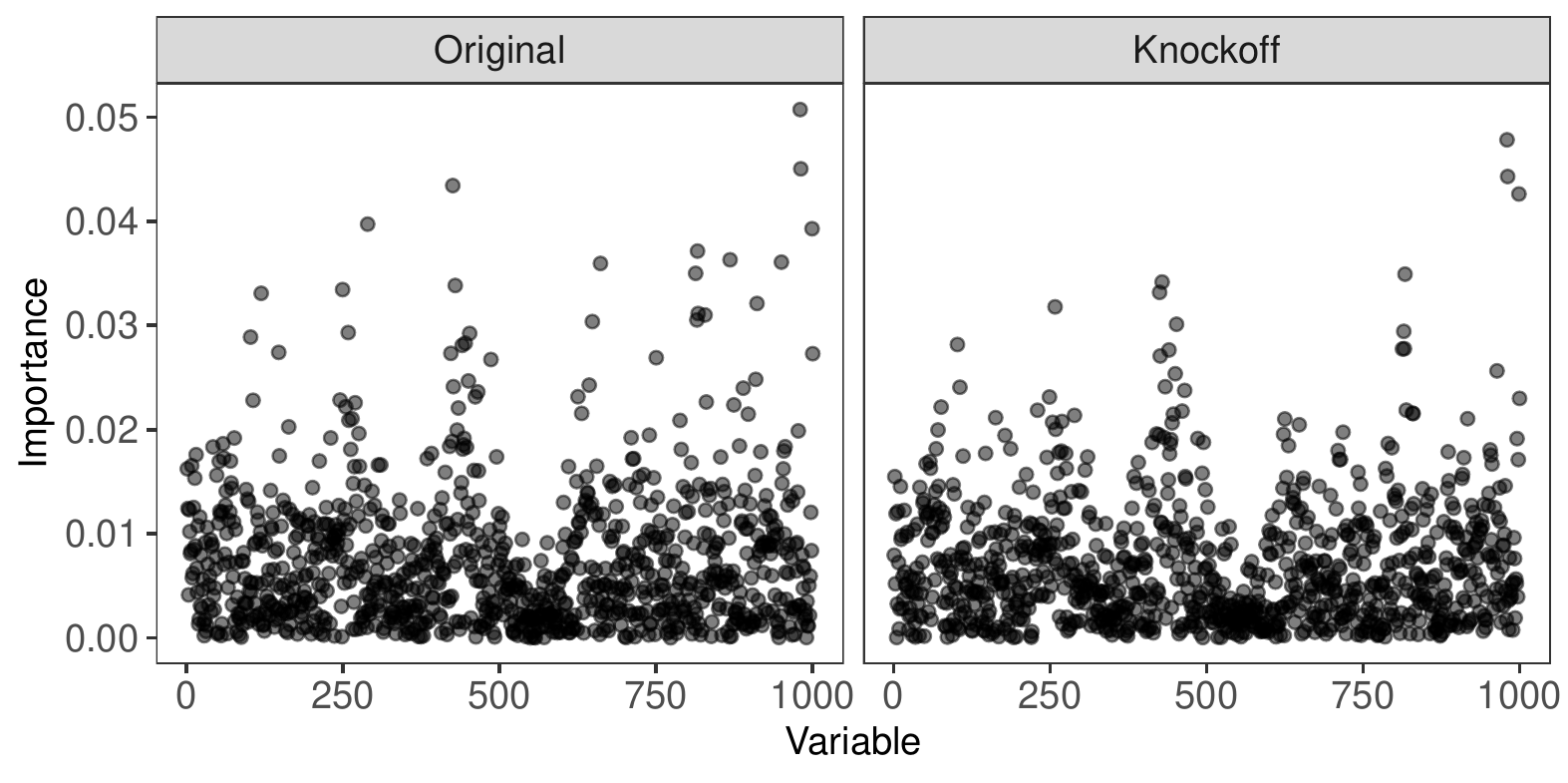}
    \caption{Knockoffs can be safely used as negative controls because they are as likely to be selected as the unimportant original variables.}  \label{fig:importance-knockoffs}
  \end{subfigure}
  \begin{subfigure}[!htb]{0.9\textwidth}
    \centering
    \includegraphics[width=0.8\textwidth]{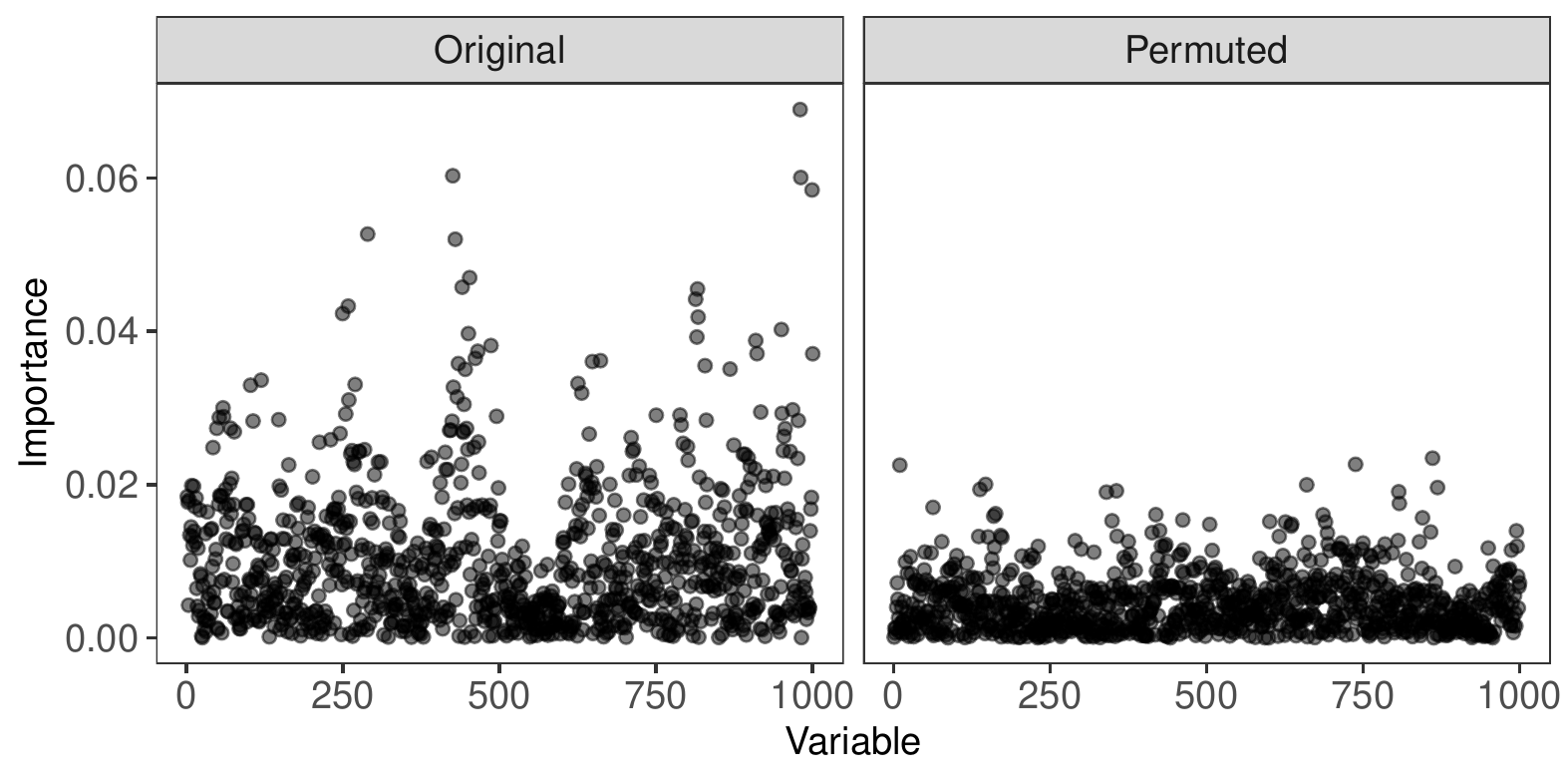}
    \caption{Permuted variables cannot be used as negative controls because they are less likely to be selected than the unimportant original variables.}  \label{fig:importance-permutations}
  \end{subfigure}
  \begin{subfigure}[!htb]{0.9\textwidth}
    \centering
    \includegraphics[width=0.8\textwidth]{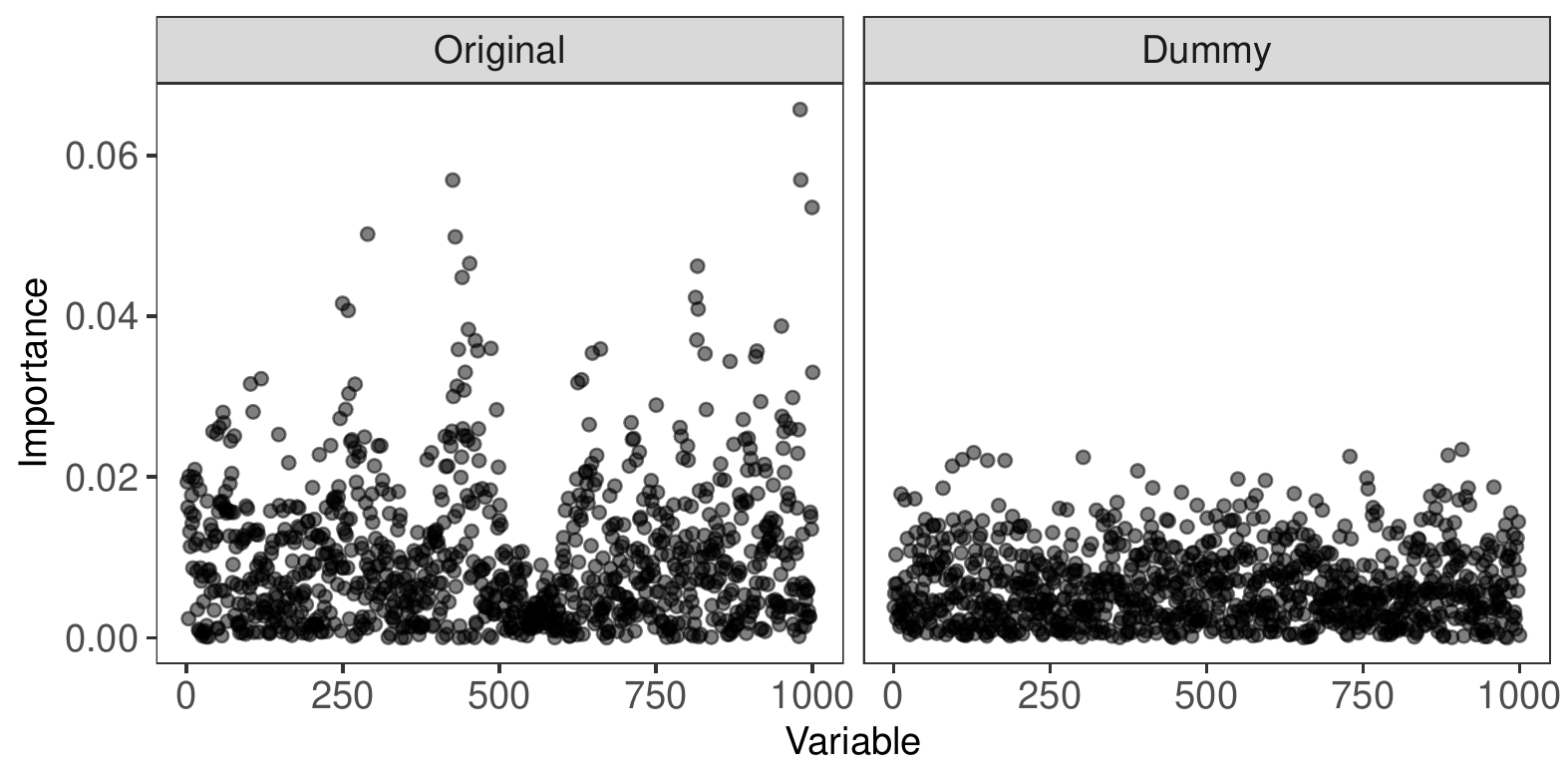}
    \caption{White-noise dummy variables cannot be used as negative controls because they are less likely to be selected than the unimportant original variables.}  \label{fig:importance-dummy}
  \end{subfigure}
  \caption{Measures of feature importance computed with ridge regression on $n=1000$ independent realizations of $p=1000$ variables sampled from a hidden Markov model and augmented with different types of artificial covariates. The response is simulated from a linear model with 60 nonzero coefficients. The 60 truly important variables, i.e., those whose coefficients in the linear model are nonzero, are not shown.} \label{fig:importance}
\end{figure}

A slightly more sophisticated example of the same principle is shown
in the numerical experiment of Figure~\ref{fig:importance}. Here, the
performance of knockoffs is compared with that of permuted variables
and independent Gaussian pseudo features, and the results provide a
striking visual representation of why such phony
variables cannot be used for calibration.

We turn to the second novelty. We live in an age where researchers
have powerful and extremely complex data fitting strategies right at
their fingertips; think of deep learning methods, sophisticated
Bayesian computations, or a combination thereof. Knockoffs are
designed to work with {\em any} feature-importance measures the
statistician would like to use---not just the time of entry in a
forward selection algorithm.

\section{Modeling the distribution of the explanatory variables}

Generating valid knockoffs requires in principle perfect knowledge of
the distribution $F_X$ of the explanatory variables. Fortunately,
  this goal is not unrealistic and a good degree of approximation can
  be achieved in practice for genome-wide association studies by leveraging the large amounts of
  available data and the prior knowledge encoded in the hidden Markov
  models of genetic variation \citep{Li2003}. It is however natural
to wonder about the behaviour of knockoffs under model
misspecification in practice.  The numerical results in Figure~2 of
\cite{jewell2019comment} are consistent with our experience that
knockoffs are typically quite robust. In fact, requiring that the
joint distribution of $\tilde{X}$ and $X$ be exchangeable in the sense
of \cite{candes2016} is much stronger than asking for false discovery rate control at a
nominal level for a specific choice of importance statistics. However,
concrete examples can be found where an incorrect sampling mechanism
leads to an inflation of the type-I errors \citep{romano2018}.

\begin{figure}[!htb]
  \centering
    \includegraphics[width=0.9\textwidth]{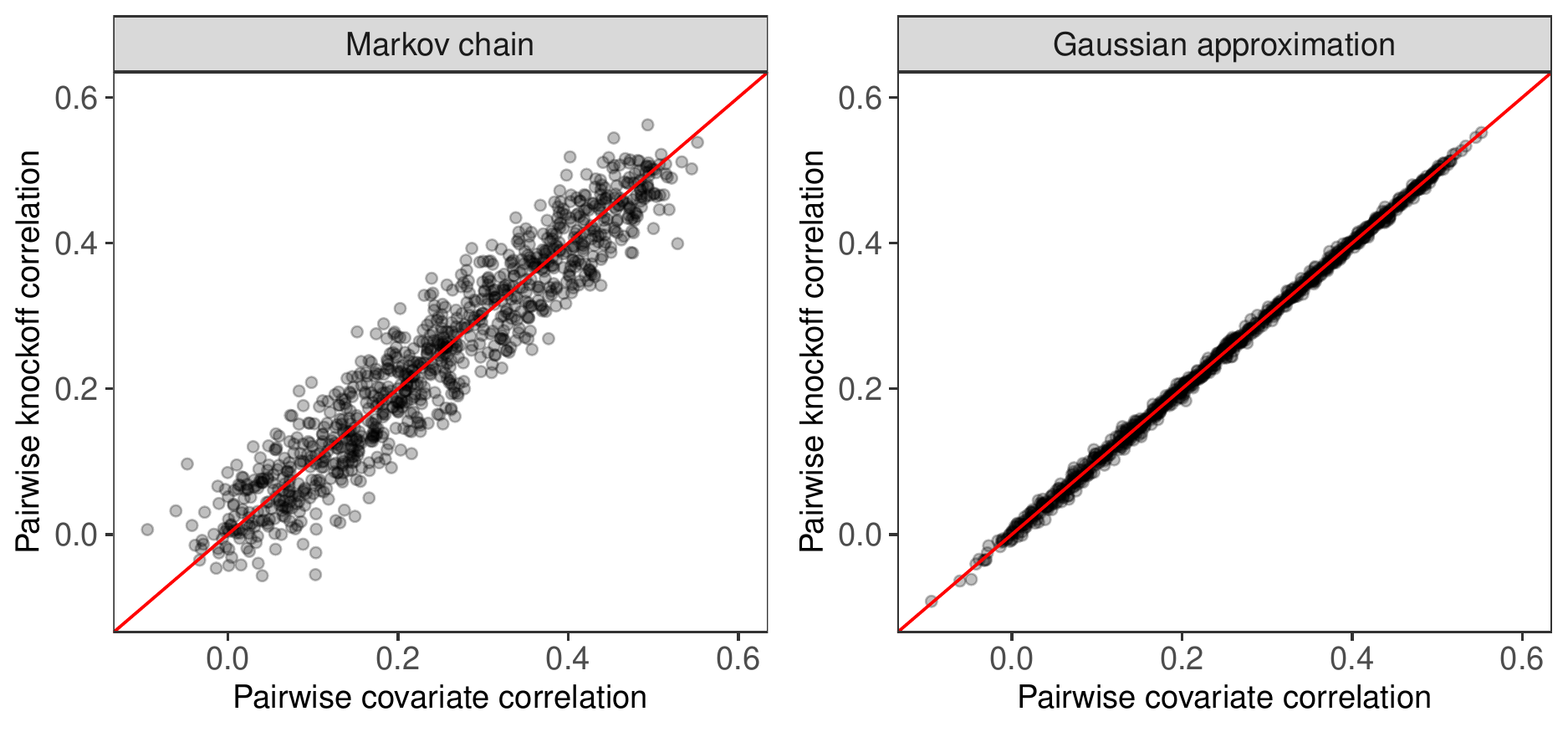}
  \caption{ Simulation with $p=1000$ covariates distributed as a Markov chain
    and exact knockoffs generated using the approach 
    in \cite{sesia2018}, as well as approximate Gaussian knockoffs obtained by applying the method of \cite{candes2016}. The experiment is the same as that of Figure 1 in \cite{jewell2019comment}, although the covariance matrix for the Gaussian knockoffs is estimated differently.
    The figure shows the pairwise correlations among the covariates, $\text{Cor}(X_j,X_{j-1})$, versus the pairwise correlations among the knockoffs, $\text{Cor}(\tilde{X}_j,\tilde{X}_{j-1})$, for $j \in \{2,\ldots,p\}$. Left: exact knockoffs for the Markov chain; right: approximate Gaussian knockoffs. Both knockoff constructions generate synthetic features whose second moments are exchangeable with the original variables. } \label{fig:diagnostics-full}
\end{figure}

\begin{figure}[!htb]
  \centering
    \includegraphics[width=0.75\textwidth]{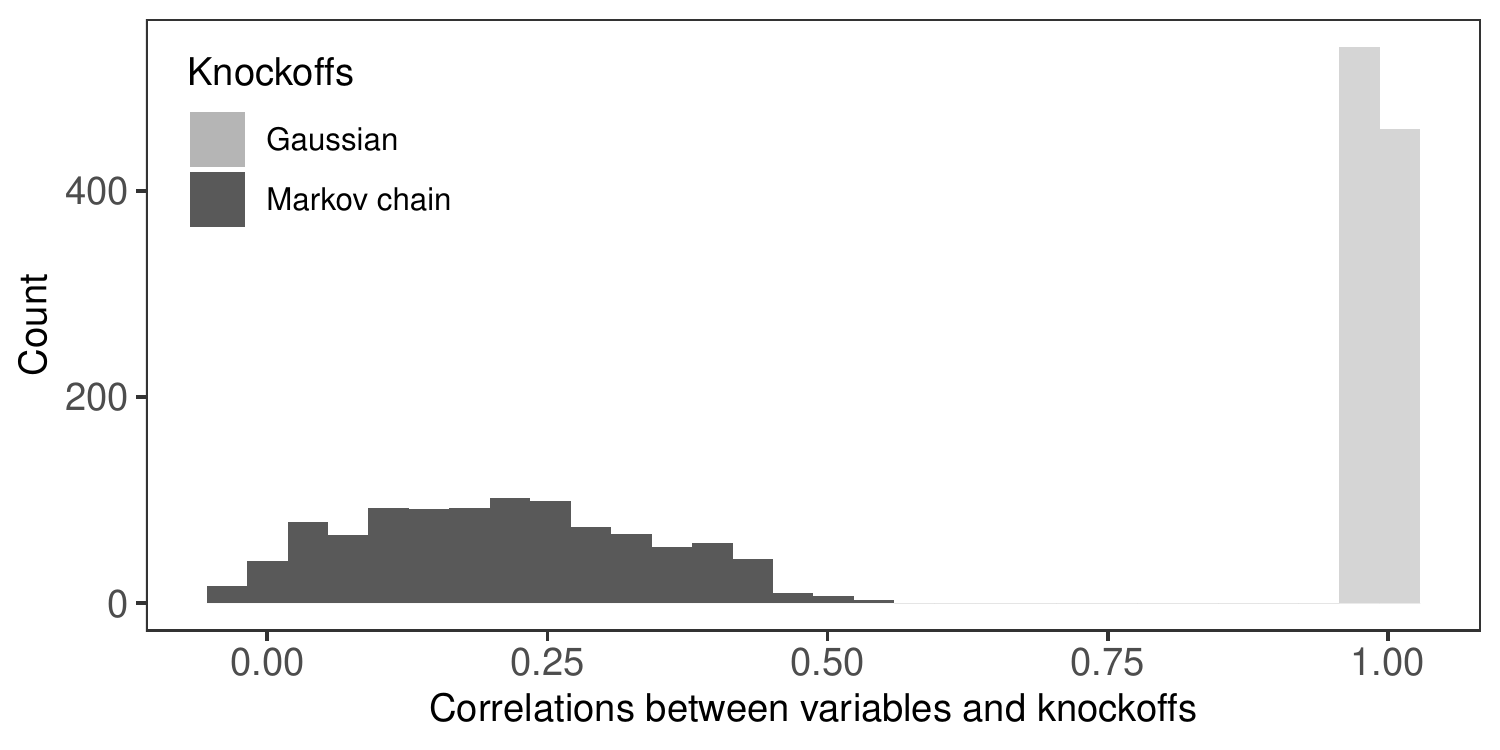}
  \caption{Histogram of the pairwise correlations between variables and knockoffs, $\text{Cor}(X_j,\tilde{X}_{j})$, for $j \in \{1,\ldots,p\}$, for the same experiment as in Figure~\ref{fig:diagnostics-full}. The approximate Gaussian knockoffs are almost identical to the original variables and will thus have very little power in practice.} \label{fig:diagnostics-self}
\end{figure}

For hidden Markov models, the approximate knockoffs of
\cite{candes2016} based on the Gaussian assumption are not rigorously
guaranteed to control the false discovery rate and they are often less powerful than
our exact construction. In order to show this point with an example,
we have replicated the experiment of Figure~1 in
\cite{jewell2019comment}, with a small technical modification, as
shown in Figures~\ref{fig:diagnostics-full} and \ref{fig:diagnostics-self}. Since the empirical covariance
matrix of $X$ is almost singular in this simulation, Gaussian
knockoffs based on the second-order approximation in \cite{candes2016}
have no power. The reason why our results in
Figure~\ref{fig:diagnostics-full} are different is that the empirical
covariance matrix was shrunk in \cite{jewell2019comment}, in the
attempt to generate non-trivial Gaussian knockoffs. On the other hand,
our algorithm provided with knowledge of the hidden Markov model
structure can generate powerful knockoffs without violating
exchangeability.

\section{Sampling knockoffs}

The conditional distribution of knockoffs $\tilde{X}$ given the
observed $X$ is not uniquely defined for any fixed data distribution
$F_X$ \citep{rosenblatt2019comment}. For example, $\tilde{X} = X$
satisfies the required exchangeability properties, despite having no
practical use. A continuous family of conditional knockoff
distributions is known for Gaussian variables, in which case one is
typically chosen by solving a semi-definite program to minimize the
pairwise correlations between $X_j$ and $\tilde{X}_j$, in order to
maximize power \citep{candes2016}. Even though a similar optimization
problem does not arise as naturally in the context of hidden Markov
models, different constructions are available. For instance, the
suggestion of \cite{rosenblatt2019comment} to set $\tilde{Z} = Z$ in
Algorithm 2 of \cite{sesia2018} would also generate exact knockoffs,
albeit more correlated with $X$. 

The recent work of \cite{romano2018} proposes an alternative machine
that can produce approximate knockoffs in great generality, without
making modeling assumptions on $F_X$ \citep{bottolo2019comment,
  rosenblatt2019comment}. The approach of \cite{romano2018} is based
on deep generative models and it can be powerful because it is driven
by the effort to make $\tilde{X}$ as uncorrelated with $X$ as possible
in the spirit of \cite{barber2015}. However, deep knockoffs are more
computationally expensive and are not exactly exchangeable when
applied to hidden Markov models. Whether ideas from our paper can be
combined with those in \cite{romano2018} to obtain an improved
knockoff sampler is an open question. In any event, deep knockoffs may
offer a practical solution for the analysis of other types of data for
which a reliable model of $F_X$ is either unavailable or intractable.

\section{Computational efficiency}

We believe that a multivariate analysis with knockoffs is in principle feasible even for very large datasets \citep{rosenblatt2019comment}, although some computational aspects of our pipeline can be improved.
The computational cost of the algorithms for sampling knockoff copies of hidden Markov models described in this paper is $\mathcal{O}(npK^2)$, where $K$ is the number of latent states. Even though this is not exorbitant compared to that of estimating $F_X$ and evaluating multivariate measures of feature importance, it can become important when $K$ is large \citep{marchini2019comment, bottolo2019comment, rosenblatt2019comment}. Moreover, all three of the major aforementioned steps in our variable selection procedure can be expensive when many samples must be considered. A solution mitigating this limitation will be presented soon, as we have developed a significantly faster implementation of our methods and we are applying it to the genetic analysis of the UK Biobank data \citep{Bycroft2018}. We will be excited to share the genotype knockoffs for this resource, as soon as we have been able to generate them with all the properties we have discussed here \citep{marchini2019comment}. 
 In any case, knockoffs are already more computationally efficient in this paper than the existing alternatives for high-dimensional conditional testing, such as the randomization test \citep{rosenblatt2019comment}, discussed in \cite{candes2016}.

\section{Aggregating dependent discoveries}


We have observed in several numerical experiments that the results
obtained from different realizations of the knockoffs can often be
combined while empirically controlling the false discovery rate; e.g., keeping only
those variables that are selected at least 50\% of the times
\citep{Gosia}. Whether more stable procedures and rigorous results can
be derived is still under investigation, since it is plausible that
such simple heuristics may fail in certain cases. We are optimistic
that future work will bring further improvements in this direction,
even though aggregating the results of different dependent tests is a
problem that goes well beyond the scope of knockoffs
\citep{jewell2019comment}. Most statistical findings are more or less
randomized, as they involve some form of data splitting, resampling,
cross-validation or simply the discretion of the practitioner to
choose a model, use prior knowledge, or tune hyperparameters.

\section{The statistical power of knockoffs}

Methods based on knockoffs may enjoy 
substantial power as a result of the
flexibility offered in the choice of the importance statistics. In
fact, a variety of sparse estimators, cross-validation techniques,
Bayesian models and very complex machine learning tools can be
summoned at will to evaluate importance measures for the augmented set
of original predictors and knockoffs. 
Of course, the choice of the most appropriate importance statistics for the problem is left to the user, who has to balance power and computational costs, and not dictated by the knockoff procedure. 
With this regard, we find the suggestion of \cite{marchini2019comment} particularly useful: it certainly seems promising to leverage the 
advantages of linear mixed model methodologies for the analysis of genome-wide association studies. For example, one can imagine using a screening procedure that is based on the results of linear mixed models on original and knockoff genotypes to obtain a smaller set of variables to pass on to a lasso estimation. We will certainly invest some effort in identifying which importance statistics are most effective and we would be thrilled to see other scientists 
contribute to this effort.

\begin{figure}[!htb]
    \centering
    \begin{subfigure}[!htb]{0.49\textwidth}
        \centering
        \includegraphics[width=0.9\textwidth]{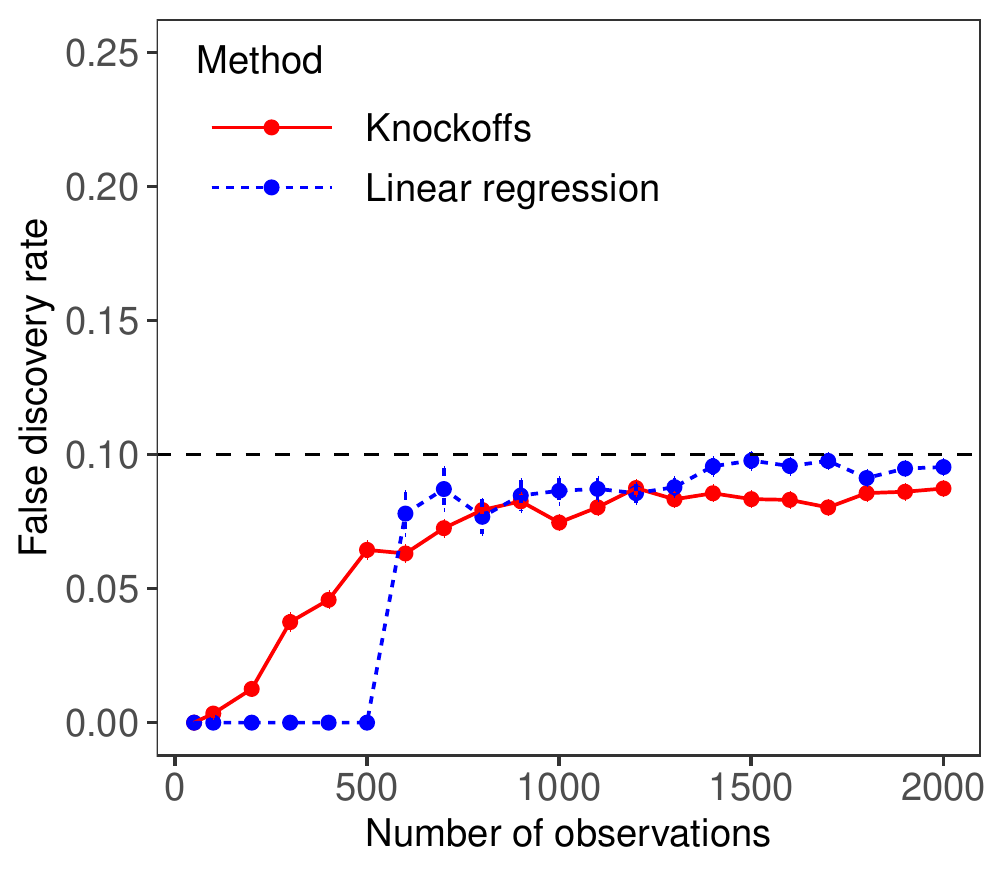}
    \end{subfigure}
    \begin{subfigure}[!htb]{0.49\textwidth}
        \centering
        \includegraphics[width=0.9\textwidth]{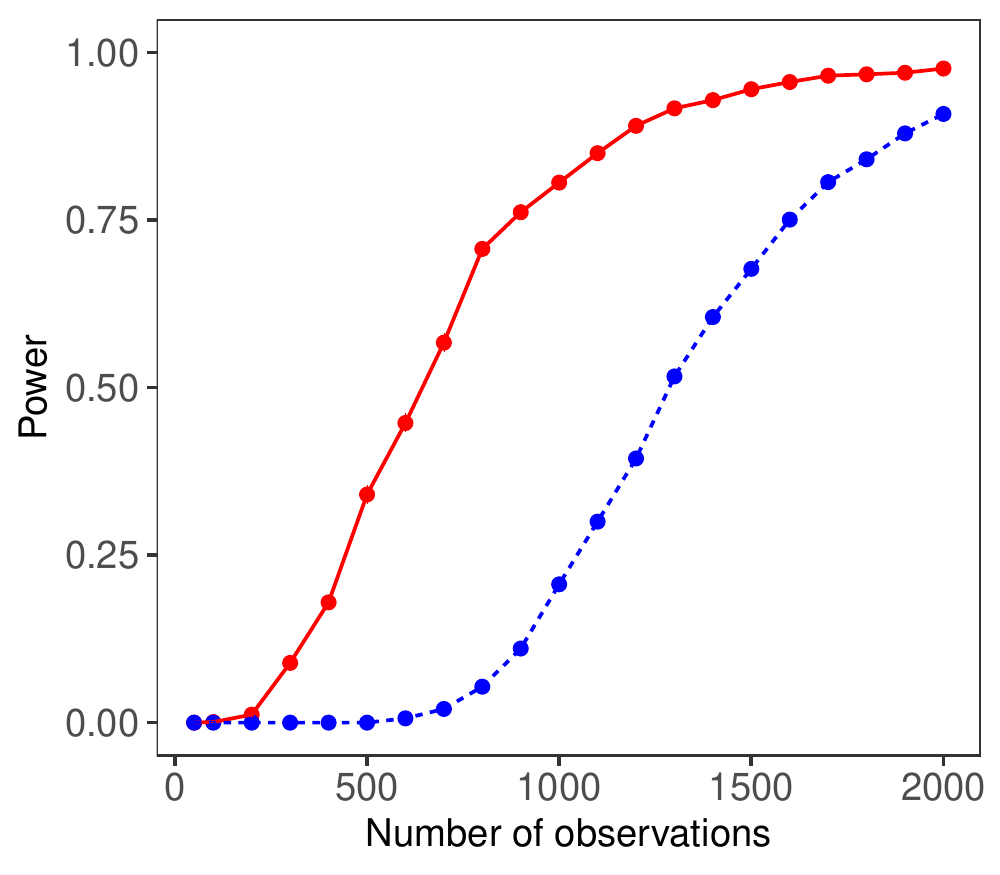}
    \end{subfigure}
  \caption{Numerical experiment comparing the performance of knockoffs and linear regression for variable selection with $p=500$ variables, in the adversarial setting designed by \cite{rosenblatt2019comment} to be unfavorable to knockoffs. The false discovery rate (left) and power (right) are averaged over 1000 independent experiments. Both methods are applied at the nominal false discovery rate level $q=0.1$, although only knockoffs are rigorously guaranteed to control it.} \label{fig:adversarial}
\end{figure}

The  argument  made by
\cite{rosenblatt2019comment} in their hypothetical example to illustrate the potential lack of power of knockoffs rests on
the assumption that knockoffs must rely on linear regression.
Although this assumption is unjustified,
knockoffs prove to be very successful even in this adversarial 
setting that they describe.
In order to test this claim, we have implemented the
experiment outlined in \cite{rosenblatt2019comment}, using $p=500$
variables divided in blocks of size 2 with internal correlations equal
to $\rho=0.9$. The number of non-null variables is
$|\mathcal{S}| = 20$ and their signal amplitude is equal to
$\beta_j=0.25$. The performance of knockoffs with Lasso statistics is
compared to that of the linear regression method suggested by
\cite{rosenblatt2019comment}, combined with the Benjamini-Hochberg
procedure \citep{benjamini1995} at the nominal level $q=0.1$. This
choice is intended to make the comparison with knockoffs as fair as
possible, although the Benjamini-Hochberg procedure is not
theoretically guaranteed to control the false discovery rate in this multivariate
regression problem.  The results reported in
Figure~\ref{fig:adversarial} show that knockoffs are much more
powerful than the proposed alternative, even though the experiment was
designed to be ``clearly unfavorable'', quoting from
\cite{rosenblatt2019comment}.

The presence of strong correlations among the covariates always makes
the variable selection problem harder, but it has no more effect on
knockoffs than it would have on any other procedure, as shown in the
experiment above. In general, knockoffs methods perform well for
essentially two reasons: they can exploit powerful measures of
variable importance and they can leverage prior information on the
structure of the predictors. We leveraged predictor information in the
experiment as $\tilde{X}$ was generated using knowledge of the
covariance of $X$.  As long as this information is available, at least
approximately, knockoffs methods can prove powerful while controlling
the false discovery rate. This justifies deployment in genome-wide association studies since a great deal of prior
information is available about the structure of the explanatory
variables.

\section{Supplementary material}

 The code to reproduce the numerical simulations described in this discussion is available online at \url{https://bitbucket.org/msesia/gene_hunting_discussion/}.

\bibliographystyle{biometrika}
\bibliography{bibliography}

\end{document}